\documentclass[
    ,final            
    ,nomathfonts
]
  {aipproc}

\layoutstyle{8x11single}
\usepackage{psfig}

\newcommand{\barr}{\begin{array}} 
\newcommand{\bea}{\begin{eqnarray}} 
\newcommand{\beq}{\begin{equation}} 
\newcommand{\ear}{\end{array}} 
\newcommand{\eea}{\end{eqnarray}} 
\newcommand{\eeq}{\end{equation}} 

\def\am{angular momentum\  } 
\def\conf{configuration\ } 
\def\confs{configurations\ }

\newcommand{\vjp}{\mbox{\boldmath $j$}_p}

\newcommand{\vjh}{\mbox{\boldmath$j$}_h}

\newcommand{\vJ}{\mbox{\boldmath$J$}}

\begin{document}

\title{Left-handed nuclei}

\author{V.\ Dimitrov }{
address={IKH, Research Center Rossendorf, PF 510119, 01314 Dresden, Germany},
}
\author{F.\ D\"onau }{
address={IKH, Research Center Rossendorf, PF 510119, 01314 Dresden, Germany},
}
\author{\underline{S.\ Frauendorf}}{
address={Department of Physics, University of Notre Dame,
Notre Dame, Indiana 46556, USA},
altaddress={IKH, Research Center Rossendorf, PF 510119, 01314 Dresden, Germany}
}

\begin{abstract}
The orientation of the \am vector with respect to the triaxial
density distribution selects a left-handed or right-handed system
principal axes. This breaking of chiral symmetry manifests itself as 
pairs of nearly identical $\Delta I=1$-bands. The chiral structures
combine high-j particles and high-j holes with a triaxial rotor.
Tilted axis cranking calculations predict the existence of such
configurations in different mass regions. There is experimental evidence
in odd-odd nuclei around mass 134. The quantized motion of the \am
vector between the left- and right-handed configurations, 
which causes the splitting between the chiral sister bands, 
can be classified as tunneling (chiral rotors) or oscillation
(chiral vibrators).  
\end{abstract}

\maketitle


\section{Introduction}
\begin{figure}[h]
\psfig{file=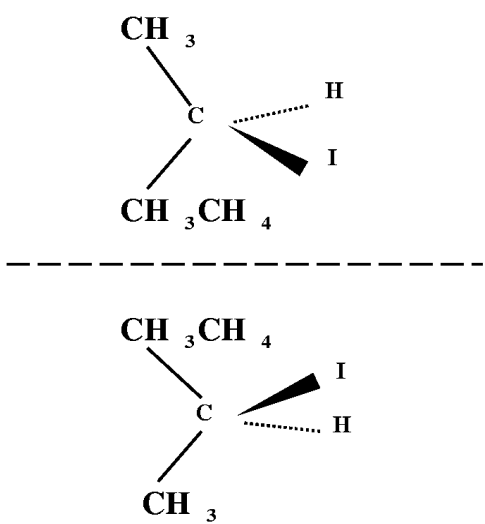,width=7cm,angle=-90}  
\caption{The two enantiomers of 2-iodubutene. The broken line 
indicates the mirror plane. }  
\label{f:chiralmol}
\end{figure}

Chirality is a common property of molecules. 
Fig. \ref{f:chiralmol} shows a simple example. The 2-iodubutene contains a
stereo center - the C atom -, to which the four different groups are
attached. If one selects the  bond to the group CH$_3$CH$_2$,
the three groups I, H, CH$_3$ form a right-handed or a left-handed screw. 
These two "enatiomers" are related to each other by mirror reflection. 
Complex bio-molecules are all chiral. Chirality is very obvious for the 
DNA double-helix. Although in principle two entantiomers exist, which
have exactly the same binding energy, organisms synthesize only one.
The reason is that the DNA blueprint provides only for one.
Since the function of bio-molecules depends critical on their geometry,
(e.g. the  key-lock mechanism of enzymes) the other enantiomer with the
mirror-geometry may not function in the organism. 
Life based on the opposite  enantiomers of all bio-molecules
 should be possible.
How came the choice about in the beginning of evolution? Most likely, it was
just by accident. However, there are speculations that the chirality
of elementary particles might have played a role.


\begin{figure}[h]
\psfig{file=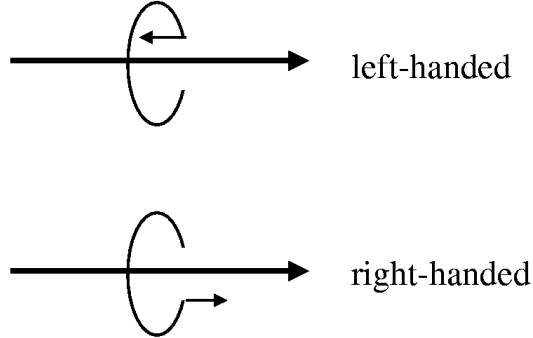,width=8cm,angle=-90}  
\caption{Chirality of a mass-less particle. The circle with 
the arrow indicates the orientation of the spin. }  
\label{f:chiralpart}
\end{figure}

Fig. \ref{f:chiralpart} illustrates chirality of a mass-less particle.
The spin can be parallel or anti-parallel to particle momentum.
Since the spin of the particle is an axial vector, it defines the a sense of
rotation and the two orientations correspond to a right-handed and
 a left-handed
system. The neutrinos, which appear only as left-handed species, 
introduce chiral asymmetry into the world. The chirality of molecules is   
of geometrical nature, the chirality of mass-less particles is dynamical.

\begin{figure}[h]
\psfig{file=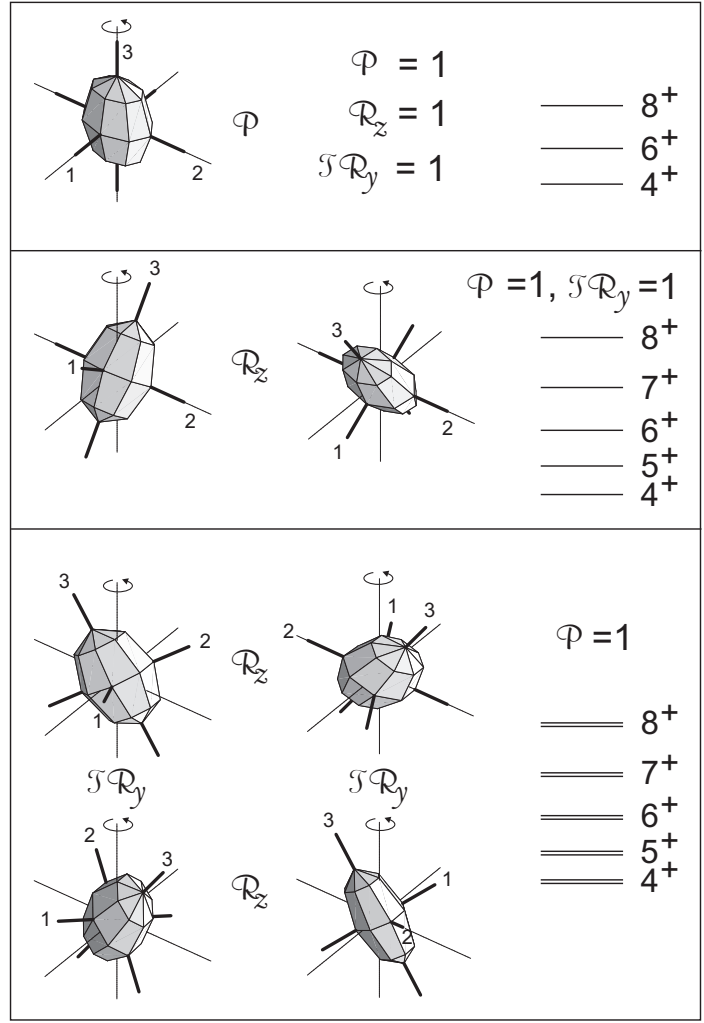,width=10cm}  
\caption{The discrete symmetries of the mean field of a rotating triaxial  
reflection symmetric nucleus (three mirror planes). The axis of rotation (z)  
is marked by the circular arrow. It coincides with the angular momentum\ $%
\vJ$. The structure of the rotational bands associated with each symmetry  
type is illustrated on the right side.The meaning of the 
symmetry operations is explained in the text.
 Note the change of chirality induced  
by ${\cal T  R}_y(\pi)$ in the lowest panel.}  
\label{f:sym}
\end{figure}

Nuclei have been thought as being achiral, because they consist of 
only two species of nucleons and have relatively
simple shapes as compared to molecules, 
However, Meng and Frauendorf \cite{chiral} recently pointed out that angular 
momentum adds a new dimension, such that the  rotation of triaxial nuclei may
attain a chiral character. The lower panel of Fig. \ref{f:sym} illustrates
this surprising possibility.  
 We denote the three principal   
axes (PA) of triaxial density  
distribution   
by l, i, and s, which stand for long, intermediate and short,  
respectively. The angular momentum\ vector $\vJ$ introduces chirality by  
selecting one of the octants. In four of the octants  
 the axes l, i, and s form a  
left-handed and in the other four a right-handed system. This gives rise to  
two degenerate rotational bands because all octants are   
energetically equivalent. Hence  
 the  chirality of nuclear rotation   
results from a combination of dynamics (the  
angular momentum) and geometry (the triaxial shape).  

\section{Tilted Axis Cranking Calculations.}

If we speak about the  shape of a nucleus, we mean the shape of its
density distribution. The symmetry of the density distribution -spherical or
deformed- decides if the spectrum will be irregular or show rotational bands.
The density distribution is found by means of mean field approaches,
like the various types of the Hartree-Fock calculations or the Strutinsky
method. For large angular momentum one has to use the Cranking 
generalizations of these methods, which describe an uniformly
rotating mean field. In these studies,  one used to assume that the axis
of uniform rotation coincides with one of the principal axes of the density
distribution, as it its the case for molecules. Frauendorf \cite{tac} 
demonstrated that nuclei are different from molecules. They may uniformly
rotate about an axis that is tilted with respect to the principal axes of
the density distribution. The Tilted Axis Cranking model (TAC) 
\cite{Kerman, Frisk,tac} 
allow us to calculate the orientation of the rotational axis. TAC 
consist in applying one of  mean field approximations to the two-body
Routhian
\begin{equation} 
H'=H-\omega J_z.
\end{equation} 
Here, $H$ is the two-body Hamiltonian of choice and $J_z$ the angular 
momentum component along the z-axis which is the rotational axis.
The new element as compared to traditional cranking calculations 
is to allow for all orientations of the density distribution with
respect to the z-axis. Fig. \ref{f:sym} illustrates how changing 
the orientation of the rotational axis leads to different discrete symmetries,
which show up in the level sequence of rotational bands.
In the  
upper panel the axis of rotation (which is chosen to be z) coincides with  
one of the PA, i. e. the finite rotation ${\cal R}_z(\pi)=1$. This symmetry  
implies the signature quantum number $\alpha$, which restricts the total  
angular momentum\ to the values $I=\alpha+2n$, with $n$ integer  
($\Delta I=2$ band) \cite{bf}.  
In the middle panel the rotational axis lies in one of the planes spanned by  
two PA (planar tilt). Since  now 
${\cal R}_z(\pi)\not=1$, there is no longer a  
restriction of the values $I$ can take. The band is a sequence of states the  
$I$ of which differ by 1 ($\Delta I =1$ band). There is a second symmetry in  
the upper two panels: The rotation 
${\cal R}_y(\pi)$ transforms the  
density into an identical position but changes the sign of the angular  
momentum\ vector $\vJ$. Since the latter is odd under the time reversal  
operation ${\cal T}$, the combination ${\cal T R}_y(\pi)=1$.

In the lower panel the axis of rotation is out of the three planes spanned  
by the PA. The operation ${\cal T R}_y(\pi)\not=1$. It changes the  
chirality of the axes l, i and s with respect to the axis of rotation $\vJ  
$. Since  the left- and the right-handed 
solutions  have a the same  
energy, they  give rise to two degenerate $\Delta I =1$ bands. They are the  
linear combinations of the left- and right-handed \confs, 
which restore  the spontaneously broken  
${\cal T R}_y(\pi)$ symmetry.

\begin{figure}[h]  
\psfig{file=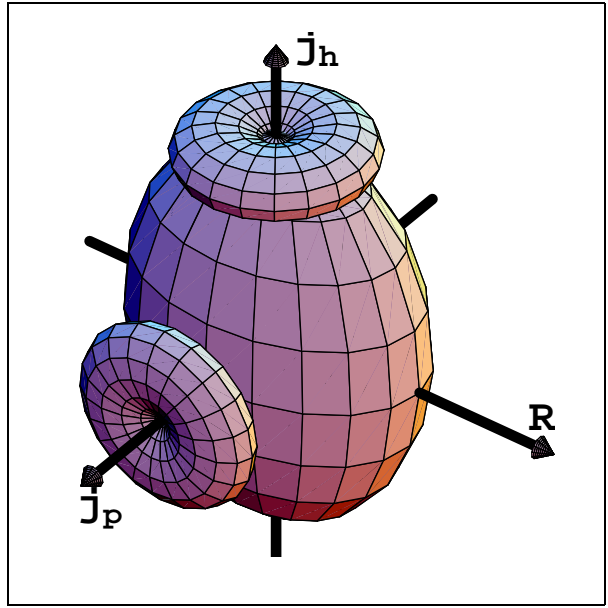,width=8cm}  
\caption{Orbitals of a high-j proton and a high-j neutron hole coupled to the
 triaxial density distribution.}  
\label{f:orb}  
\end{figure}  

Fig. \ref{f:orb} illustrates how such a solution may arise. The proton aligns
its angular momentum $\vjp$ with the short axis of the density
 distribution. This orientation maximizes the overlap of its orbital 
with the triaxial density, which corresponds to minimal energy, because
 the core-particle interaction is attractive.   The neutron hole aligns
its angular momentum $\vjh$ with the long axis. 
This orientation minimizes the overlap of its orbital 
with the triaxial density, which corresponds to minimal energy, because
 the core-hole interaction is repulsive. The angular momentum of the core
$\vec R$ is of collective nature. It likes to 
orient along the intermediate axis,
which has the largest moment of inertia, because the density distribution 
deviates strongest from rotational symmetry with respect to this axis.

\begin{figure}[h]
\mbox{\psfig{file=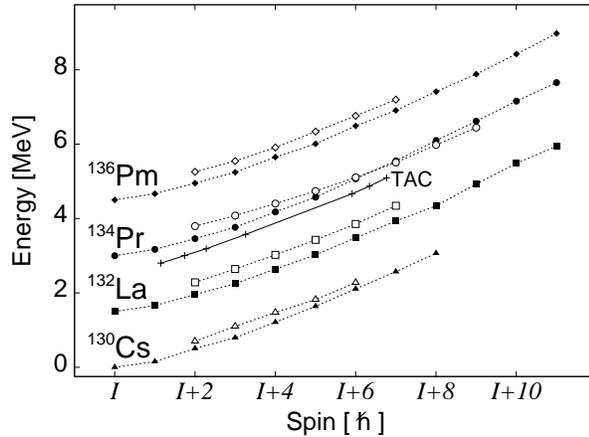,width=8cm,angle=-90}}
\hspace*{3cm}
\caption{\label{f:expsys}Chiral sister bands in the $N=75$ isotones
based on the configuration $\pi h_{11/2} \nu h_{11/2}^{-1}$.
 The parity of the bands is + and $I=9$. From \protect\cite{chiralsys}.
The TAC calculation from \protect\cite{chiralprl} is included.
}
\end{figure}

The chiral configuration is rather special. The shape must be triaxial,
and there must be high-j particles and high-j holes.
The existence of such a structures and their location in the nuclear
chart must by studied by means of microscopic TAC calculations.
The status of these explorations will be discussed in the  last section.
Dimitrov, Frauendorf and D\"onau \cite{chiralprl}
found the first completely self-consistent
chiral solution for $_{59}^{134}$Pr$_{75}$ with the  maximal triaxiality
of $\gamma\approx 30^o$. Fig. \ref{f:expsys} shows that in this nucleus 
two $\Delta I =1$ bands, which are based on the
$[\pi h_{11/2},\nu h_{11/2}^{-1}]$ configuration, merge forming
 a  doublet structure \cite{pr134}. Consistent with the experiment, the 
TAC solution attains  chiral character only for $I>15$.  

Pairs of close bands with the same parity are found in the neighboring 
odd-odd nuclei as well \cite{chiralsys,bark,starosta,hecht,hartley}.
Fig. \ref{f:expsys} shows as examples the isotones 
of$_{59}^{134}$Pr$_{75}$. The bands are close but do not merge. 
This means that the chirality cannot be broken
in a static way as in molecules. There must be substantial tunneling
between  the left- and right handed configurations. As we discuss in the 
next section, the a weaker form of chirality may occur 
as a slow motion of  
the angular momentum vector into the left- and right handed sectors.
The frequency of such a chiral vibration is less than 300 keV
(cf. Fig. \ref{f:expsys}). In this respect, breaking of chiral symmetry 
is similar to breaking reflection symmetry by octupole deformation.
The two branches of opposite parity do not completely merge into
one rotational band (see e.g. \cite{octupole}.

For most molecules tunneling between the two enatiomers is negligible;
for complex molecules the tunneling time
 easily exceeds the age of the universe.
We can separate the left-handed from the right-handed
enantiomers (or other organisms do it for us) and can experiment 
with one specie. The turning of the polarization plane of light by 
organic sugar is a familiar example. The effect is caused by the
chiral arrangement of the optical active atoms in the molecule.
In nuclei, chirality is harder to demonstrate, because of the rapid motion
between the left- and right-handed configurations. Therefore, one has to 
understand this motion. A microscopic description of this large amplitude
motion is not yet available for the interpretation of the experiment. 
However, we may gain insight into the dynamics by further
studying the particle-rotor model, in the context of which the possibility
that rotating triaxial nuclei may become chiral was conceived \cite{chiral}.

\section{Dynamics of Chirality in the Particle-Rotor Model}

The simplest chiral configuration is illustrated in Fig. \ref{f:orb}.
One high-j proton and one high-j neutron couple to a triaxial rotor.
For the odd-odd nuclei in the mass 134 region, the high-j orbitals are 
$h_{11/2}$.
The generic case of a 
triaxial rotor with maximal  
asymmetry ($\gamma =30^{o}$) and the irrotational flow  
relation ${\cal J}_{l}={\cal J}_{s}={\cal J}_{i}/4$ between 
the moments of inertia was studied in \cite{chiral,bark,starosta}. 

\begin{figure}[h]
\mbox{\psfig{file=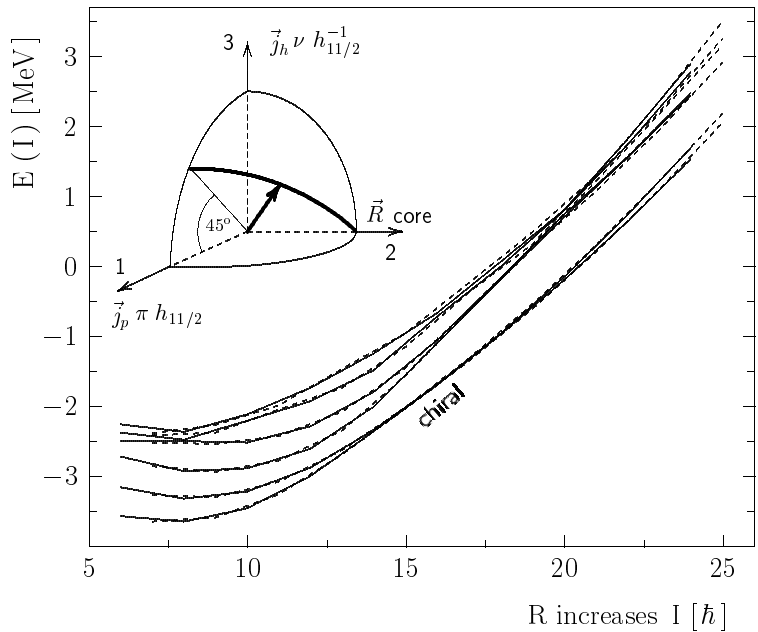,width=10cm}}
\caption{\label{f:pr}
Rotational levels of  $h_{11/2}$ particles and holes coupled to
a triaxial rotor with $\gamma=30^o$.
 Full lines: $\alpha=0$ (even $I$). 
Dashed lines to $\alpha=1$ (odd $I$). The insets show the orientation of the
angular momentum with respect to  the triaxial potential, where 
1, 2 and 3 correspond
to the short, intermediate and long  principal axes, respectively.
The angular momentum vector moves along the heavy arc. 
The  position displayed 
corresponds to the spin interval $13 <I<18$, where the two lowest bands
are nearly degenerate.  The right-handed position is shown.
The left-handed is obtained by reflection through the 1-3 plane.
From \protect\cite{chiral}.  
}\end{figure}
 
Fig. \ref{f:pr} shows the result of such calculations. 
At  the   beginning 
 of the lowest band the angular momentum originates from the 
particle and the hole,  
whose individual angular momenta are aligned with the s- and l-axes. These  
orientations correspond to a maximal overlap of the particle and hole  
densities with the triaxial potential, as illustrated in Fig. \ref{f:orb}. 
A $\Delta I=1$ band is generated by  
adding the rotor momentum $\vec{R}$ in the s-l plane (planar tilt). There is  
a second $\Delta I=1$ band representing a vibration of $\vec{J}$ out of the  
s-l plane, which is generated by a wobbling of $\vec{R}$. This is 
a more precise description of the chiral vibration mentioned above. 
Higher in the band, $\vec{R}$ reorients toward the i-axis, which  
has the maximal moment of inertia. The left- and the right-handed positions  
of $\vec{J}$ separate. Since they couple only by some tunneling,  
the two bands come very close together  in energy.
This is the regime we called chiral rotation.
 The reorientation of $\vec{R}$, i. e. the transition from chiral vibration to 
rotation is well localized in the spectrum Fig. \ref{f:pr}. It appears also
in the higher bands at larger $I$. 

\begin{figure}[t]
\psfig{file=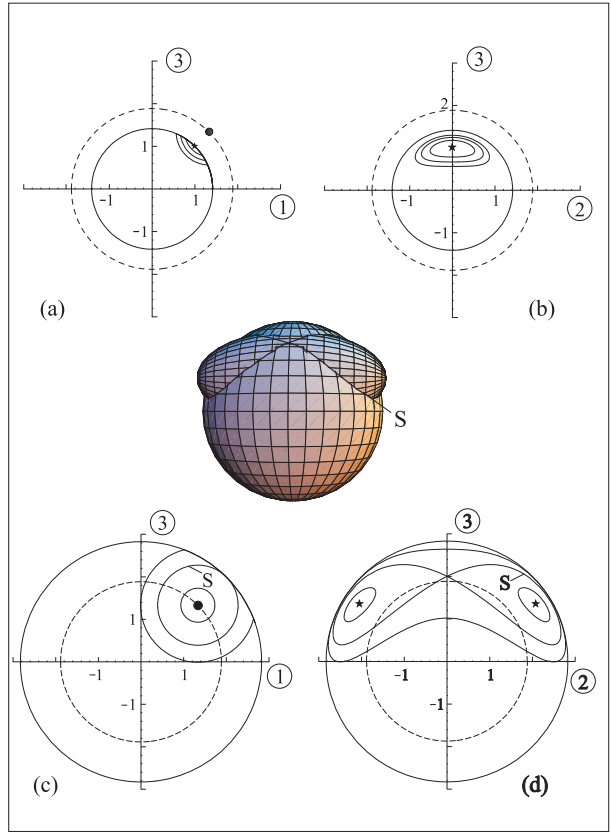,width=8cm}
\vspace{0.5cm}
\caption{\label{f:sep} 
Classical orbits of the angular  momentum of triaxial rotor coupled to 
a particle and a hole with spin $j$. The central figure shows a 
three-dimensional illustration of the intersecting surfaces of the 
angular momentum  sphere and
energy spheroid. The classical orbit is located at the intersection.
The other figures show projections of the orbits on the 1-3 plane (a, c)
and 2-3 plane (b,d). The dashed circle shows the critical angular 
$\bar{J}_{crit}=4/3\sqrt{2}$. Figs. (a) and (b) 
correspond to $\bar{J}=2/3\sqrt{2}$ and (c) and (d) to $\bar{J}=6/3\sqrt{2}$.
All angular momenta are in units of the particle angular momentum $j$.  
The separatrix is denoted by $S$. The discrete orbits are only
illustrative. They  do not fulfill the quantization condition.  }
\end{figure}

A more detailed picture of this transition emerges if we consider 
the classical orbits of the angular momentum vector $\vJ$.   
For a  fixed value $J$, the endpoint of  $\vJ$ is confined 
to the surface of the sphere 
\begin{equation}\label{J}
J^2=J_1^2+J_2^2+J_3^2.
\end{equation}
The energy of the triaxial rotor is given by 
\begin{equation}\label{E}
E=\frac{\hbar^2}{2{\mathcal J}_i}\,[4(J_1-j)^2+J_2^2+4(J_3-j)^2 \,]\,.
\label{Eclass}
\end{equation}
where 1, 2, 3 denote the s-, i- and l-axes, respectively.
Further we assume that $\vjp$ and $\vjh$ have the same magnitude and
 are parallel to the s- and l-axes (cf. Fig. \ref{f:orb}), 
and that they remain frozen in this orientation. In this 
``frozen alignment'' approximation, the energy of the particle-rotor system, 
which is conserved, is given by (\ref{E}) (up to a 
constant). Equation (\ref{E}) defines
an spheroidal energy surface on which the angular momentum
 vector has to end. The classical 
orbits of $\vJ$ are given as the  intersections of 
the \am sphere and the energy spheroid.    
 In Fig.\,\ref{f:sep} a typical set of classical  
orbits is plotted, where we introduce the 
dimensionless quantities $\bar{ \vJ}=\vJ /j$ and 
$\bar{ E}=E \times 2{\mathcal J}_i/(\hbar j)^2$. Increasing the
energy of the states for given $J$,  
corresponds to increasing size of the spheroid.  The 
symmetry axis of the spheroid is parallel to the 2-axis at  
$\bar{J}_1=\bar{J}_3=4/3$.  Its position defines a critical \am.
If $\bar J < 4\sqrt{2}/3$, the axis lies outside the angular momentum
sphere. As seen in the upper panel, the lowest energy correspond to the
touching point of the sphere and spheroid, which is marked by the star.  
The lowest orbits revolve this point in the 1-3 plane. They are
the chiral vibrations, which are seen as  the lowest three bands in
Fig. \ref{f:pr} for $J<10$. The name alludes to the fact that
the vector $\vJ$ oscillates between the left- and right-handed sectors.  

If $\bar J > 4\sqrt{2}/3$, the axis lies inside the angular momentum
sphere. As seen in the lower panel of Fig. \ref{f:sep},
at the minimal energy, the energy spheroid touches 
the \am sphere from inside at the {\it two} symmetric points marked 
by  the stars.
For somewhat larger energy, the spheroid penetrates the \am sphere,
which gives rise to two separate orbits around the two points 
of minimal energy. This case corresponds to static chirality or  
chiral rotation, where all three \am components are non-zero.
Fig. \ref{f:pr} shows a calculation for $h_{11/2}$ particles, i.e.
$j=5.5$ and the critical \am $4\sqrt{2}/3\times 5.5=10.3$. Above this
value, the two lowest bands quickly merge into a chiral doublet.
The quantal tunneling smooths out the transition from the chiral vibrational
to the chiral rotational regime.    
When $J$ is sufficiently larger than the critical \am, there 
 there are more than one orbit around the minima possible,
 which correspond to the excited doublets of bands in Fig. \ref{f:pr}.
These states are characterized by a wobbling motion around the
two points of minimal energy.
At still larger energy the front of the spheroid comes out of the sphere.
The left-handed and right-handed orbits are no longer separated, i.e. we are
back to the regime of only dynamical chirality 
without the characteristic doublets of bands. 
The energy of this transition corresponds
to a special orbit, the separatrix, which is shown in 
the center of Fig. \ref{f:sep}.

The static energy barrier between the left- and right-handed sectors is 
in the order of 100 $keV$ for $J\approx 15$ (cf. Fig.3 in \cite{chiral}). 
This seems very low. Nevertheless, the quantal calculation shows that
the splitting between the bands is very small (cf. Fig. \ref{f:pr}) and that 
the wavefunction  is concentrated around the minima of the classical energy. 
Tunneling is determined by {\it the ratio
of the barrier height and the mass coefficient}. The mass coefficient 
is small enough such that the tunneling through the low barrier is strongly
reduced.

\section{Predictions of chirality in various mass regions}\label{s:sys}

Chirality becomes possible if high-j particles and high-j holes couple to
a triaxial core. We have 
carried out TAC calculations in order to locate
mass regions, where these conditions are met. Table 1 summarizes the
results. The table shows only one representative nucleus for each region.
Most TAC calculations have been carried out for the
region around $_{59}^{134}$Pr$_{75}$, which we 
discuss now. One can expect similar behavior
in the regions around the other representative nuclei.

\begin{table}[h]
\begin{minipage}[b]{7cm}
\mbox{\begin{tabular}{ccccccc}
\hline
$Z$	&$N$	&particle	&hole	&$J$	&$\varepsilon$	&$\gamma$\\
\hline
35	&44	&$\pi g_{9/2}$	&$\nu g_{9/2}$	&9	&0.19	&26\\
43	&65	&$\nu h_{11/2}$	&$\pi g_{9/2}$	&13	&0.21	&14\\
59	&75	&$\pi h_{11/2}$	&$\nu h_{11/2}$	&13	&0.18	&26\\
\hline
\end{tabular}}
\caption{
Representative nuclei for which TAC
 calculations give chiral 
solutions. The particle and hole orbitals, which
align with the short and long
axes, are indicated.
The column $J$ displays the \am with the
strongest
 chirality. The last two columns contain the shape
 parameters for this value of $J$. }
\end{minipage}
\hspace*{0.5cm}
\hfill
\begin{minipage}[b]{7cm}
\mbox{\begin{tabular}{ccccccc}
\hline
$Z$	&$N$	&particle	&hole	&$J$	&$\varepsilon$	&$\gamma$\\
\hline
77	&111	&$\pi i_{13/2}$	&$\nu i_{13/2}$	&13	&0.21	&40\\
69	&93	&$\pi i_{13/2}$	&$\nu h_{11/2}$	&45	&0.32	&26\\
\hline
\end{tabular}}
\end{minipage}
\end{table}

Both the experiment and the TAC calculations point to an island of chirality
around $Z=59$ and $N=75$. In the $N=75$ chain, the shores seem to be 
$Z=65$ and 51. The low-$Z$ shoreline seems to be 72. The high-$Z$ shore is 
not known yet. The center of the island is  $Z\approx60$ and 
$N\approx 76$.  
The TAC-barriers between the the left- and right handed sectors are 
in the order  of 50- 100 $keV$. The Particle-Rotor calculations show that
this leads to grouping into the chiral sister bands. 
However, for a quantitative 
estimate of the splitting one needs calculating the tunneling in
a microscopic way, which we have not been able yet.  
Therfore, we only correlate the chiral TAC solutions
with the appearance of pairs of bands, without trying to predict if
they will be chiral vibrators or rotors.     

The simplest chiral configurations appear in odd-odd nuclei, where
a high-j particle and high-j hole couple to the triaxial rotor. Most of the
experimental chiral sister bands in the mass 134 region have this structure.
The positive parity orbits do not contribute strongly to the \am. 
Combining the chiral high-j structure  with one particle on such a spectator 
orbit gives a  chiral \conf in an odd-A nucleus. As an example, TAC
 calculations predict  chiral sister bands with the \conf 
$[\pi (dp)h_{11/2}, \nu h_{11/2}^{-1}]$ in $^{135}_{60}$Nd$_{75}$. Adding
another spectator embeds the chiral skeleton  into an even-even  nucleus.
TAC calculations give the chiral \conf
 $[\pi (dg)h_{11/2}, \nu(dg) h_{11/2}^{-1}]$ in $^{136}_{60}$Nd$_{76}$.
For breaking the chiral symmetry, the particle and hole ``legs'' need not be
of the same length, and they may be composed of more than one high-j orbital.
As an example, we calculated the chiral 
\conf $[\pi h_{11/2}^2,\nu h_{11/2}^{-1}]$ in  
$^{135}_{60}$Nd$_{75}$. Zhu et al. \cite{nd135} found a pair of negative
parity bands coming as close as 60 $keV$ in this nucleus, which have 
this structure. Mergel et al. found
a pair of negative parity bands in $^{136}_{60}$Nd$_{76}$, 
which may be   $[\pi h_{11/2}^2,\nu (dg)h_{11/2}^{-1}]$ or, as
suggested by the authors, 
 $[\pi (dg)h_{11/2},\nu h_{11/2}^{-2}]$. 
  
\begin{table}[h]
\begin{tabular}{ccccccccccccc}
\hline
$Z \backslash N$&67	&68	&69	&70	&71	&72	&73	&74	
&75	&76	&77	&78\\
\hline
65		&	&	&	&	&	&	&	&	&	&	&	&\\
64		&	&	&	&	&	&	&	&	&	&	&	&\\
63		&	&	&	&	&	&	&	&	&S/CV	&	&	&\\
62		&	&	&	&	&	&	&	&	&	&	&	&\\
61		&	&	&	&	&	&	&	&	&L/CR	&	&	&\\
60		&	&	&	&	&	&	&	&	&L/CR	&L/CR	&L/0	&\\
59		&	&	&P/AC	&	&	&	&0/CV	&	&L/CR	&	&0/CV	&\\
58		&	&	&	&	&	&	&	&	&	&L/0	&	&\\
57		&	&	&	&	&	&	&0/CV	&	&M/CV	&	&0/CV	&\\
56		&	&	&	&	&	&P/AC	&	&	&	&	&	&\\
55		&	&	&	&	&	&	&0/CV	&	&S/CV	&	&0CV	&\\
54		&	&	&	&	&	&	&	&	&	&	&	&\\
\hline
\end{tabular}
\caption{\label{t:sys} Chirality  in the mass 134 region. TAC calculations: P - {\bf p}lanar solution, S - chiral solution in 
a {\bf s}hort $J$-intervall, and L - chiral solution in 
a {\bf l}ong $J$-interval. Observation of chiral doubling: CR - {\bf c}hiral {\bf r}otation (the two $\Delta I =1$
bands come very close), CV - {\bf c}hiral {\bf v}ibration (the two $\Delta I =1$
bands remain separated), AC - {\bf ac}hiral (no evidence for pairs of bands). 
The 0 indicates no calculation or insufficient experimental
information. Experiments are taken from 
\protect\cite{chiralsys,bark,starosta,hecht,
nd136,nd135}.  The calculations use the hybrid version of TAC as described in 
\protect\cite{chiralprl}, where some results are published.   }		
\end{table}

\begin{figure}[h]
\psfig{file=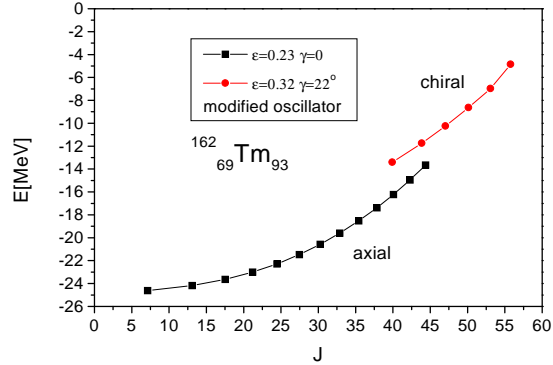,width=8cm}  
\caption{TAC prediction of a chiral band in $_{69}^{162}$Tm$_{93}$
for highly deformed triaxial shape.
No pairing is taken into account.
The yrast band with normal deformation has even spin and positive parity.
 The normal-deformation  bands with other parity and odd spin are near yrast.
 The TAC calculations
use the modified oscillator potential.  }  
\label{f:tm162}
\end{figure}

Much less TAC calculations have been carried out for the other regions 
in Table 2. The only experimental candidate for chiral sister bands 
is $^{104}_{45}$Rh$_{59}$ \cite{rh104}. 
Strongly deformed triaxial shapes at high spin
have been suggested for nuclei with $Z=70,71$ and $N=93-96$
where evidence for the wobbling excitations has been found \cite{wobbler}. 
The strongly deformed triaxial \confs contain the $i_{13/2}$
proton particle. For the lowest  neutron numbers the 
$h_{11/2}$ hole becomes accessible and chirality a possibility. 
TAC calculations predict the existence of such a structure in
$^{162}_{69}$Tm$_{93}$. Fig. \ref{f:tm162} shows that around spin 50 the 
chiral sister bands should approach the yrast line.


\section{Conclusions}

The quantal rotation of molecules has been the paradigm for the interpretation
 of nuclear rotational bands. However, for nuclei  
the relation between angular momentum $\vJ$ and
velocity \mbox{\boldmath{$\omega$}}
 is much more complex than for  molecules. 
Nuclei contain nucleons on orbits with large 
angular momentum, which
is kept constant by quantization. Due to the presence of these 
micro-gyroscopes
the  axis for uniform rotation (the
angular momentum vector \vJ)  can take any 
direction with respect to the  density distribution.
The 
chiral symmetry is broken if $\vJ$ does not lie in one of the three mirror 
planes planes of the triaxial density distribution. 
Chirality appears in geometric arrangement
of atoms or chemical groups in in molecules. 
Chirality is also a dynamical property of mass-less elementary particles.
The new example of chirality of rotating nuclei results
from the combination  geometrical (triaxial shape) and dynamical (orientation
of \vJ) features. 

Chiral rotation manifest itself as a pair of 
nearly identical $\Delta I=1$-bands
with the same parity. Tunneling between the left- and right-handed 
configurations causes an energy splitting between the chiral sister bands.
A weaker form chirality are the chiral vibrations, 
which are slow oscillations of \vJ \,between the left-
and right-handed configurations. They show up as an equidistant
sequence of $\Delta I=1$-bands, separated by the (small) vibrational
energy. 

Microscopic tilted axis cranking calculations predict chirality 
in various mass regions. There is some systematic experimental evidence
for chiral sister bands in odd-odd nuclei around mass 134, which is one of
the predicted regions.     


\begin{theacknowledgments}
This work was supported in part by the U. S. Department of Energy, Nuclear
Physics Division, Grant DE-FG02-95ER40934.
\end{theacknowledgments}


\bibliographystyle{aipprocl} 

\end{document}